\documentclass[epj,nopacs]{svjour}
\usepackage{graphics}
\begin{document}
\hyphenation{Gos-kon-t-ract}

\title
      {New analysis on narrow baryon resonance decaying into $pK^0_s$
	in $pA$-interactions at $70\ GeV/c$ with SVD-2 setup.}
\authorrunning{A.~Aleev et al.}
\titlerunning{New analysis on narrow baryon resonance with SVD-2 setup.}
\author{A.~Aleev\inst{3} \and
E.~Ardashev\inst{2} \and
V.~Balandin\inst{3} \and
S.~Basiladze\inst{1} \and
S.~Berezhnev\inst{1} \and
G.~Bogdanova\inst{1} \and
I.~Boguslavsky\inst{3} \and
V.~Bychkov\inst{3} \and
V.~Ejov\inst{1} \and
G.~Ermakov\inst{1} \and
P.~Ermolov\inst{1} \and
N.~Furmanec\inst{3} \and
V.~Golovkin\inst{2} \and
S.~Golovnia\inst{2} \and
S.~Gorokhov\inst{2} \and
I.~Gramenitsky\inst{3} \and
N.~Grishin\inst{1} \and
Ya.~Grishkevich\inst{1} \and
D.~Karmanov\inst{1} \and
A.~Kholodenko\inst{2} \and
A.~Kiriakov\inst{2} \and
N.~Kouzmine\inst{3} \and
V.~Kozlov\inst{1} \and
E.~Kokoulina\inst{3} \and
V.~Kramarenko\inst{1} \and
A.~Kubarovsky\inst{1} \and
I.~Kudryashov\inst{1} \and
V.~Kuzmin\inst{1} \and
E.~Kuznetsov\inst{1} \and
G.~Lanshikov\inst{3} \and
A.~Larichev\inst{1} \and
A.~Leflat\inst{1} \and
M.~Levitsky\inst{2} \and
S.~Lyutov\inst{1} \and
M.~Merkin\inst{1} \and
A.~Minaenko\inst{2} \and
G.~Mitrofanov\inst{2} \and
V.~Nikitin\inst{3} \and
S.~Orfanitsky\inst{1} \and
V.~Parakhin\inst{2} \and
V.~Petrov\inst{2} \and
V.~Peshekhonov\inst{3} \and
A.~Pleskach\inst{2} \and
V.~Popov\inst{1} \and
V.~Riadovikov\inst{2} \and
V.~Ronjin\inst{2} \and
I.~Rufanov\inst{3} \and
D.~Savrina\inst{1} \and
V.~Senko\inst{2} \and
N.~Shalanda\inst{2} \and
M.~Soldatov\inst{2} \and
L.~Tikhonova\inst{1} \and
T.~Topuria\inst{3} \and
Yu.~Tsyupa\inst{2} \and
A.~Uzbyakova\inst{1} \and
M.~Vasiliev\inst{2} \and
A.~Vishnevskaya\inst{1} \and
K.~Viriasov\inst{3} \and
V.~Volkov\inst{1} \and
A.~Vorobiev\inst{2} \and
A.~Voronin\inst{1} \and
V.~Yakimchuk\inst{2} \and
A.~Yukaev\inst{3} \and
L.~Zakamsky\inst{2} \and
V.~Zapolsky\inst{2} \and
N.~Zhidkov\inst{3} \and
D.~Zotkin\inst{1} \and
S.~Zotkin\inst{1} \and
E.~Zverev \inst{1}}

\institute{Skobeltsyn Institute of Nuclear Physics,
Lomonosov Moscow State University, 1/2 Leninskie gory, Moscow, 119992 Russia
\and
Institute for High-Energy Physics,
Protvino, Moscow region, 142284, Russia
\and
Joint Institute for Nuclear Research,
Dubna, Moscow region, 141980, Russia}

\date{\today}

\abstract{
The inclusive reaction $p A \rightarrow pK^0_s + X$ was studied at
IHEP accelerator with $70\ GeV/c$ proton beam using SVD-2 detector.  Two
different samples of $K^0_s$, statistically independent and belonging
to different phase space regions, were used in the analyses and a
narrow baryon resonance with the mass $M=1523\pm 2(stat.)\pm
3(syst.)\ MeV/c^2$ was observed in both samples of the data. The
combined statistical significance was estimated to be of 8.0 (392
signal over 1990 background events). Using the part of events
reconstructed with better accuracy the width of resonance was
constrained to $\Gamma < 14~MeV/c^2$ at 95\% C.L. The $x_F$ distribution
was found to have a peak at zero with \mbox{$<|x_F|> \approx 0.1$},
that qualitatively agrees to a Regge-based model predictions. A new
cross section estimate of $\sigma \cdot BR(\Theta^+ \rightarrow pK^0) =
4.9 \pm 1.0(stat.) \pm 1.5(syst.)~ \mu b/nucleon$ for $x_F > 0$ was
obtained.
}

\maketitle
\section {Introduction.}
In the last years the observation of a narrow baryon state named
$\Theta^+$ predicted by Diakonov, Petrov and Polyakov\cite{dpp} has
been reported by a large number of experiments in the  $nK^+$ or
$pK^0_s$ decay channels
\cite{nakano,diana,clas1,saphir,itep,clas2,hermes,svd4,zeus,nomad}.
However, several experiments, mostly at high energies, did not confirm the
existence of $\Theta^+$. The complete list of references to positive
and negative results with discussion can be found in
\cite{dzierba,hicks,danilov}. The situation became more
intriguing when CLAS collaboration reported negative results
on $\Theta^+$ photoproduction off proton and deutron with high
statistics\cite{burkert}. Meanwhile LEPS collaboration reported on another
evidence of $\Theta^+$-baryon in the reaction $\gamma d\rightarrow
\Theta^+ \Lambda^*(1520)$\cite{nakano3}. The STAR collaboration
observed the double charged exotic baryon in the $pK^+$ decay
channel\cite{star}. Therefore, pentaquark existence is still under the
question and new experiments are needed to confirm or refute it.

The SVD-2 collaboration has reported earlier the observation of narrow baryon
resonance in the $pK^0_s$-system with the mass of
$M=1526\pm3(stat.)\pm 3(syst.)~MeV/c^2$ and $\Gamma <
24~MeV/c^2$\cite{svd4}.  In that analysis the $K^0_s$-mesons decayed
inside vertex detector were used (decay region 2 -- 35 mm, zero stands for
the center of the 1st target plane).  In this
paper we present our new results of the study of the same
reaction:\\ $pN\rightarrow \Theta^+ + X$, \ $\Theta^+ \rightarrow
pK^0_s$, \ $K^0_s \rightarrow \pi^+\pi^-$.\\
\noindent
We have used two independent data samples, selected by the point of
$K^0_s$ decay: inside and outside the vertex detector ( decay regions
0.2 -- 35 and 35 -- 600 mm, respectively). First results of this work
were published earlier (\cite{pq2005,svdichep06}).

\begin{center}
  \begin{figure}[ht]
    \vspace{75mm} {\includegraphics{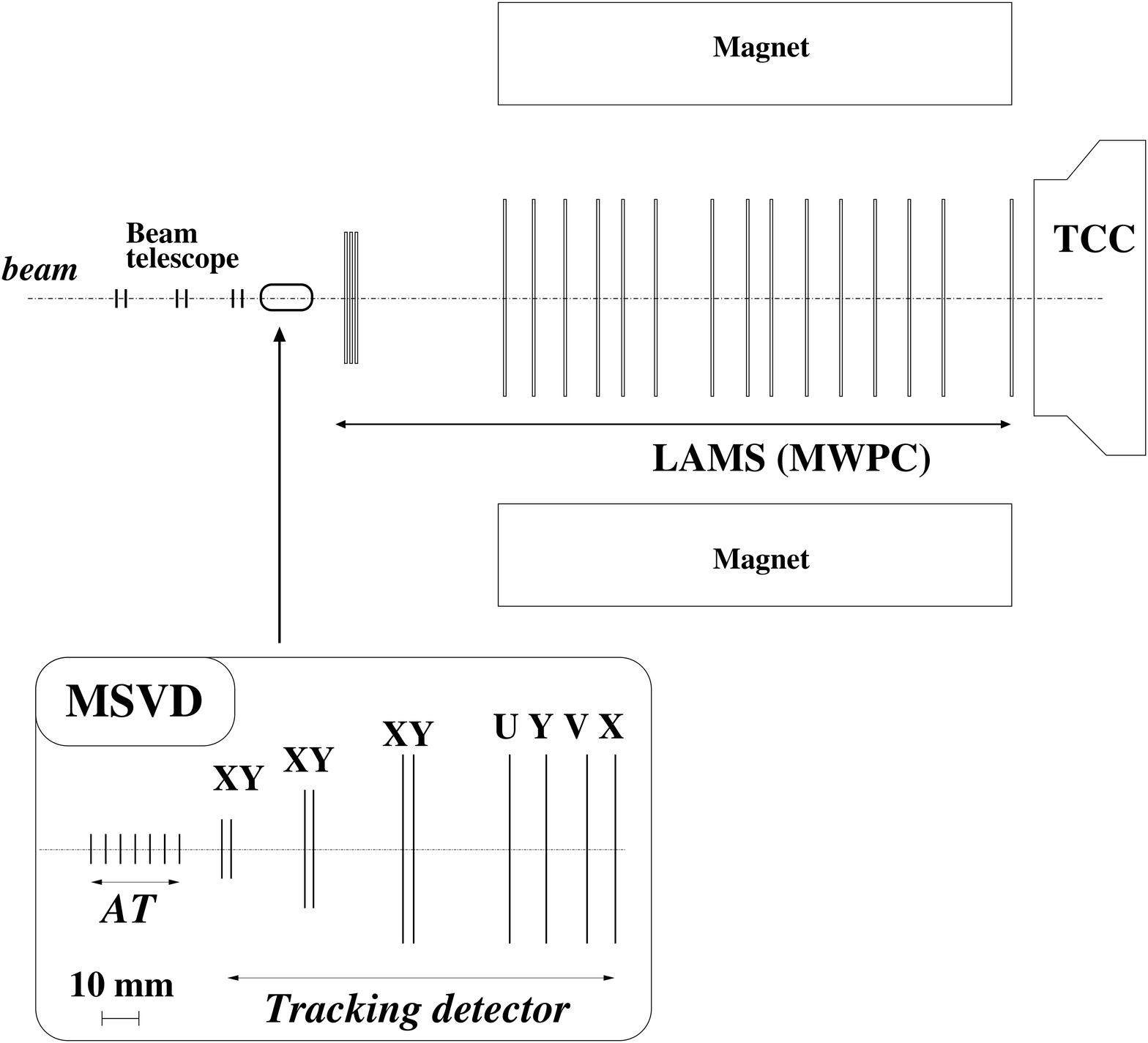}}
    \caption{SVD-2 layout (described in the text).}
    \label{setup}
  \end{figure}
\end{center}

\section {SVD-2 experimental setup.}

A detailed description of SVD-2 detector and the trigger system can be
found elsewhere\cite{svd4,vtxd,svdtrig}. For these studies, we used the
following setup components
(Fig.\ref{setup}):

\begin{enumerate}
\item the high-precision microstrip vertex detector (MSVD) with
  five active (Si) and two passive (C,Pb) target planes (AT);
\item large aperture magnetic spectrometer (LAMS) : multiwire proportional
  chambers (MWPC) with Y(vertical), U($Y-10.5^\circ$) and V($Y+10.5^\circ$)
  wire directions;
\item the multicell threshold Cherenkov counter (TCC), with the
$\pi / K /p$ thresholds of 3/11/20 GeV/c.
\end{enumerate}

Data taking was performed in the 70 GeV/c proton beam of IHEP accelerator
with an intensity of $\approx (5 \div 6) \cdot 10^5$
protons/cycle. The total statistics of $5\cdot10^7$ inelastic events was
obtained. The sensivity of this experiment for inelastic
$pN$-interactions, taking into account the trigger efficiency, was
$2.3~nb^{-1}$.

\section {Sample I: $K^0_s$ decaying inside the vertex detector
  (decay region 0.2 - 35 mm).}

After the first SVD-2 pentaquark publication\cite{svd4} (January, 2004) new
improved algorithms of the tracks reconstruction had been
developed. It increased 3-4 times the available number of $K^0_s$ decays
inside the vertex detector and essentialy improved the
experimental $K^0_s$ mass resolution.
The selected events were having a well
defined secondary vertex in the region of 0.2 -- 35 mm along the beam
direction from the beginning of the active target (corresponding to
the sensitive area of the vertex detector). Secondary tracks
were traced to the magnetic spectrometer to obtain their momenta.
About 3$\cdot$10$^4$ kaon candidates
in the events with the number of primary charged tracks
 $N_{ch}< 8~$ were selected, to reduce the combinatorial background
in further analysis.
Primary particles were traced to the spectrometer and
two-body invariant mass distributions were built using previously
selected neutral particles in association with primary vertex tracks.
$K^0_s\pi^+$ and $\Lambda \pi^+$ invariant mass spectra showed that the
masses and widths of K$^{*+}(892)$-meson and $\Sigma^{+}(1385)$-baryon
are consistent with PDG tables\cite{pdg}.

\begin{figure}[ht]
\vspace{80mm} {\includegraphics{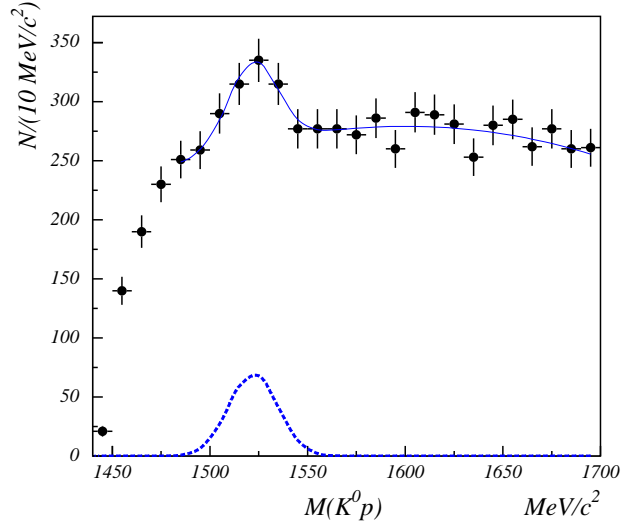}}
\caption{Sample I: The $(pK^0_s)$ invariant mass spectrum for
  the events with $N_{charged}< 8$. The dashed line stands for the peak
  extracted from the fit.}
\label{vfig4}
\end{figure}

The $pK^0_s$ invariant mass spectrum is shown in Fig.\ref{vfig4}.
An excess of s=205 signal over b=1050 background events is observed
in the area of interest, with the significance of
s/$\sqrt{s + b}$ = 5.8 .

\begin{figure}[ht]
\vspace{80mm} {\includegraphics{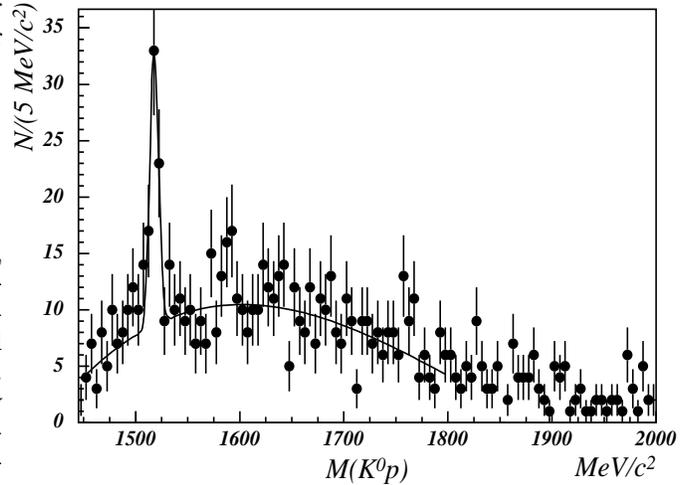}}
\caption{Sample I: The $(pK^0_s)$ invariant mass spectrum for
  $K^0_s$ decaying inside the vertex detector with additional quality
  cuts explained in text.}
\label{vfig5}
\end{figure}

To estimate the natural width of the observed peak we selected events
with the best reconstruction accuracy, applying the track quality cuts:

\begin{itemize}
  \item
    $D < 2\sigma$, \\where D is the distance of the closest approach
    for $V^0$ tracks and $\sigma$ is a calculated error,
  \item
    $N_{hits}\geq 12$, \\where $N_{hits}$ is the number of hits for
    the track in the spectrometer planes.
\end{itemize}
With that, we obtained the mass resolutions
$3.98~MeV/c^2$ for $K^0_s$ and $1.46~MeV/c^2$ for $\Lambda$ particles.

The background was reduced further with the selection of events with
$3~GeV/c < P_{proton} < 10~GeV/c$ and $N_{ch}> 3$; the latter cut may
correspond to the greater inelasticity of the events with the
$\Theta^+~$ creation \cite{Azimov}.
The result is shown in Fig.\ref{vfig5}. Here the curve
stands for the normalized mixed-event background built of kaon and
proton taken from different events with the same kinematic
requirements. The distribution was fitted by the sum of Gaussian
function and second-order polynomial background.
The Gaussian sigma of $4.1 \pm 1.0~MeV/c^2$ is consistent with
our estimate of the highest possible experimental resolution of the
SVD-2 setup. Taking into account the experimental SVD-2 resolution
(calculated using well-known resonances) it was estimated that
the intrinsic width of the $(pK^0_s)$-resonance is $\Gamma < 14~MeV/c^2$
at 95\% C.L. No signal was observed in the combinations of $K^0_s~$ with a
tracks of negative charge.

\section {Sample II: "Distant" $K^0_s$ decaying outside the vertex detector
(decay region  $35 - 600\ mm$).}

To create a Sample II, $K^0_s$-mesons and $\Lambda$-hyperons were
reconstructed by the following procedure. The events with
$n_{ch}\le 8$ in the primary vertex were reconstructed using the
algorithm described in \cite{svd4} and the spectrometer hits belonging
to the reconstructed tracks were removed. The remaining spectrometer
hits were used to search for tracks originating
from secondary vertices in the region before the first spectrometer
plane (decay region $35 - 600\ mm$). The initial track candidate had
to have at least 3 hits in the Y-planes which could be approximated by a
polynomial function originating from the area of interest. For these
candidates the hits in the U and V planes of the spectrometer were
searched for and global track parameters were defined using magnetic
field map. Tracks of opposite signs were combined to test on
having a common secondary vertex. Based on the
intrinsic tracking resolution of the spectrometer the minimum distance
between two tracks was required to be less than 1 mm in the horizontal
and 5 mm in the vertical directions. The efficiency of this method was
estimated as 50\%, it's lower than that for the reconstruction based on
vertex detector information.

\begin{figure}[ht]
\vspace{100mm} {\includegraphics{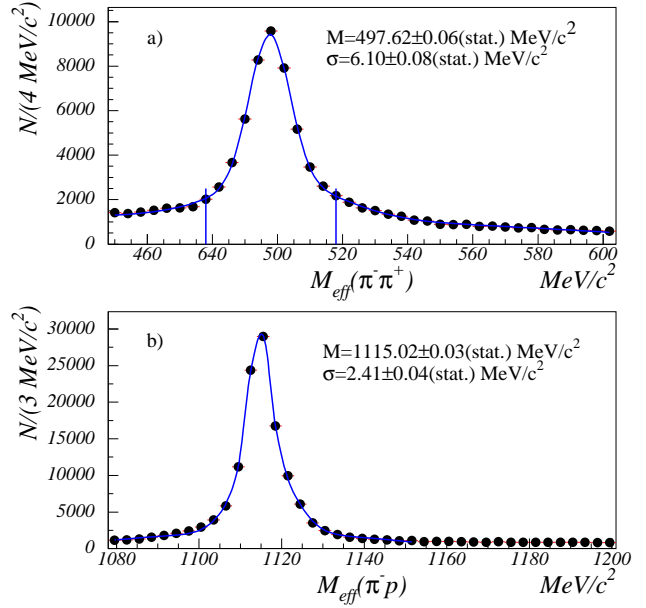}}
\caption{Sample II: a). The $(\pi^+\pi^-)$ invariant mass
  spectrum. A window corresponding to $\pm 3\sigma$ is shown by the
  vertical lines. b) The $(p \pi^-)$ invariant mass spectrum.}
\label{k0lam}
\end{figure}

\begin{figure}[ht]
\vspace{90mm} {\includegraphics{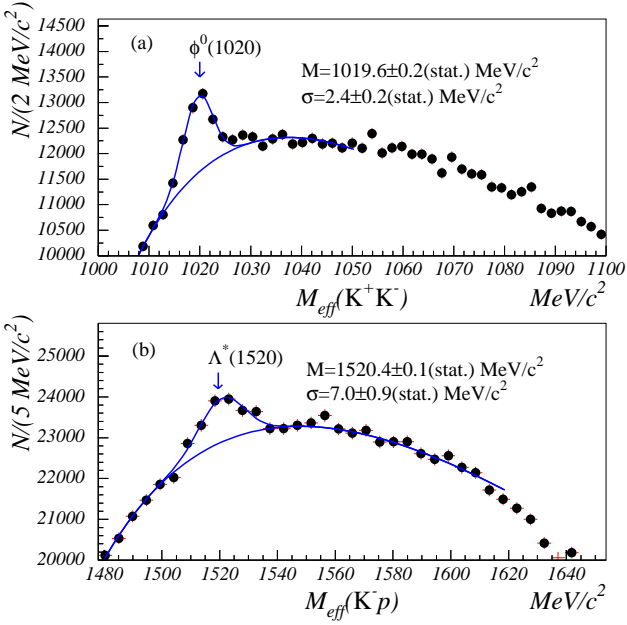}}
\caption{Sample II: a) The $(K^+ K^-)$ invariant mass spectrum. b)
  The $(K^- p)$ invariant mass spectrum.}
\label{phi}
\end{figure}

The resulting invariant masses of ($\pi^+ \pi^-$) and ($p \pi^-$)
combinations are shown in Fig.\ref{k0lam}a and \ref{k0lam}b. The
standard deviations in the mass distributions are $6.1~MeV/c^2$ and
$2.4~MeV/c^2$ for $K^0_s$ and $\Lambda$ masses respectively.
The obtained tracks collection was used as an opportunity to check for
the production of well-known resonances.
For $\phi^0(1020) \rightarrow K^+ K^-$ and
$\Lambda^{*}(1520) \rightarrow K^- p$ (Fig.\ref{phi}) a TCC particle
identification was used.
The total number of detected $\Lambda^{*}(1520) \rightarrow K^- p\/$ decays
was about $2\cdot10^4$.
Using PDG data, we estimated our experimental
mass resolutions as $\sigma_{\Lambda^{*}(1520)} = 2.4\pm0.9~MeV/c^2$ and
$\sigma_{\phi^0(1020)} = 1.6\pm0.2~MeV/c^2$.

\begin{figure}[h]
\vspace{90mm} {\includegraphics{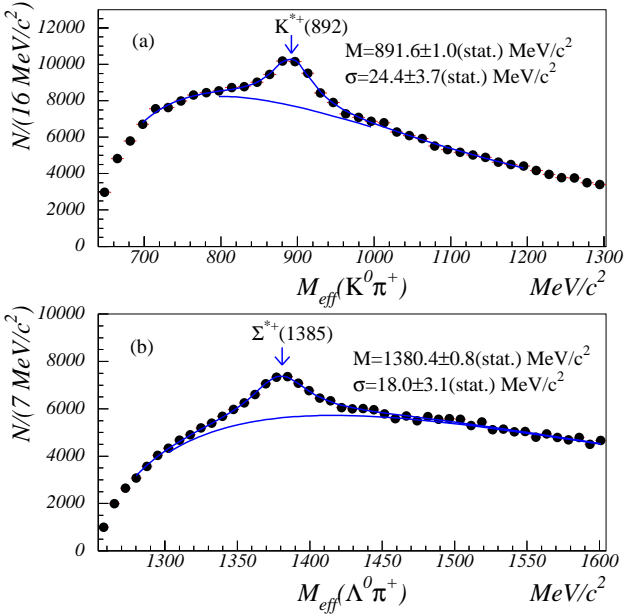}}
\caption{Sample II: a) The $(\pi^+K^0_s)$ invariant mass spectrum.
  b) $(\Lambda \pi^+)$ invariant mass spectrum. }
\label{kstar}
\end{figure}

\begin{figure}[ht]
\vspace{95mm} {\includegraphics{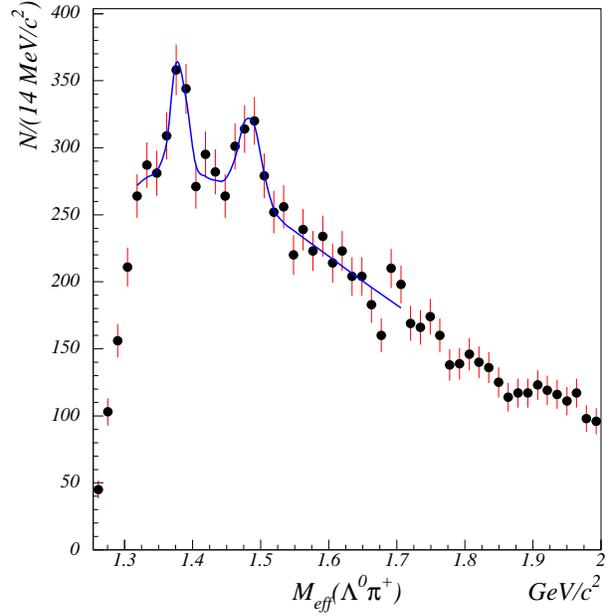}}
\caption{ The $(\Lambda \pi^+)$ invariant mass
  spectrum with the additional $p_{\Lambda}<6~GeV/c$ cut.}
\label{fig10_1}
\end{figure}

The production of the resonances decaying into strange neutral
particles was also studied.  The $(\pi^+K^0_s)$ and $(\Lambda \pi^+)$
invariant mass spectra are shown in Fig.\ref{kstar}.
The clear peaks from $K^*(892)$ and $\Sigma^+(1385)$-resonances are seen.
In total,
the masses and widths of well-known resonances
were all in a good consistency with their PDG values (\cite{pdg}).
For a $\Sigma^+(1385)$, when applying a $p_{\Lambda}<6~GeV/c$ cut,
we observed a structure near
$1480~MeV/c^2$ (Fig.\ref{fig10_1}). This peak may correspond to the
$\Sigma(1480)$, marked as one star resonance in the PDG
review(\cite{pdg}).

To search for the $\Theta^+\/$-particle, the events with multiplicity
of six or less charged tracks in the primary vertex were selected to
minimize the combinatorial background.

About $52000$ $K^0_s$-mesons in the mass window of
$\pm 20\ MeV/c^2$
were found in the selected events.
To eliminate contamination from $\Lambda$ decays,
candidates producing the effective mass of less than $1.12\ GeV/c^2$ for
($p \pi^-$) mass hypothesis were rejected. The resulting invariant mass
distribution is shown in Fig.\ref{k0lam}a.

Proton candidates were selected as positively charged tracks with a
number of spectrometer hits more than 12 and momentum $8\ GeV/c \le
P_p \le 15\ GeV/c$. Pions of such energies should leave a hit in the
Threshold Cherenkov Counter, so the absence of hits in TCC was also
required. Moreover, for the events {\em with} a TCC hit of a positive
particle taken as a proton, no significant peaks were found in the
$(pK^0_s)$ invariant mass spectra. We estimated that the
$\pi^+$-background averages to no more than $10\%$ under selection
criteria used.

\begin{figure}[ht]
\vspace{80mm} {\includegraphics{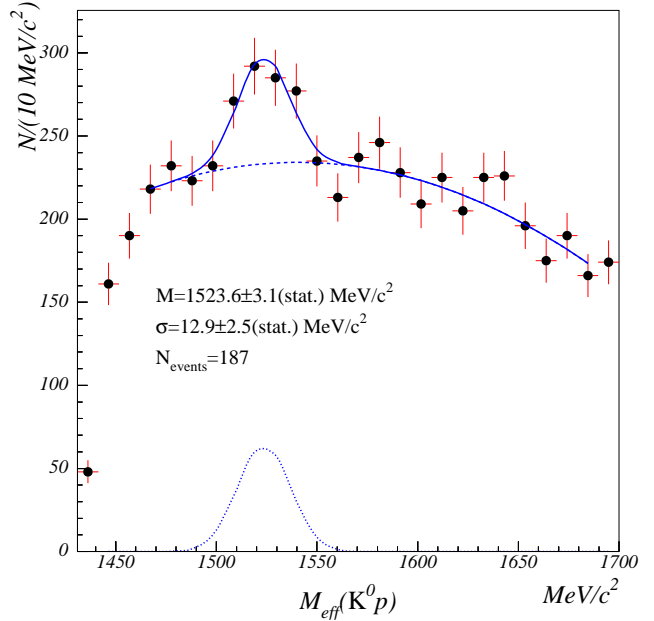}}
\caption{Sample II: The $(pK^0_s)$ invariant mass spectrum with the
  cuts explained in text.The dotted line in the bottom stands for the
peak extracted from the fit.}
\label{theta}
\end{figure}

Effective mass of the $pK^0_s$ system is plotted in
Fig.\ref{theta}. A peak is seen at the mass $M=1523.6\pm
3.1\ MeV/c^2$ with a $\sigma=12.9\pm 2.5\ MeV/c^2$. The distribution
was fitted by a sum of a Gaussian function and second-order polynomial
background. There are s=187 events in the peak over b=940 background
events. The statistical significance for the peak is
s/$\sqrt{s + b}$ = 5.6

\noindent

\begin{figure}[ht]
\vspace{95mm} {\includegraphics{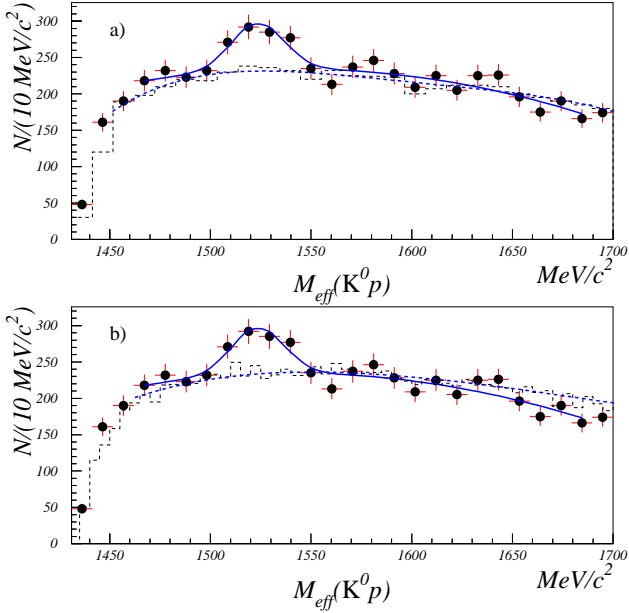}}
\caption{Sample II: The $(pK^0_s)$ invariant mass spectrum
  with different background descriptions represented by fitted dashed
  histogram: a) mixed-event model background; b) RQMD Monte Carlo
  background.}
\label{background}
\end{figure}

Two different models were tried to describe the background. The first was
taken from RQMD Monte Carlo (Fig.\ref{background}a) \cite{rqmd} and the second
from the mixed event approach (Fig.\ref{background}b). Both of them fitted
plausibly the experimental data.

It was verified that observed $pK^0_s$-resonance can not be a
reflection from other (for example $K^{*\pm}(892)~$ or $\Delta^0$)
resonances. The mechanism of producing spurious peak around
$1.54~MeV/c^2$ involving $K^0_s$ and $\Lambda$ decays overlap as
suggested by some authors \cite{zavertyaev,longo} was also checked and
no ghost tracks were found.

\section {Monte-Carlo simulation of pA-interactions.}

The invariant mass spectra of different particles and $K^0_s$ in pA
collisions have been simulated by means of RQMD 2.3 event
generator\cite{rqmd}.

\begin{figure}[ht]
\vspace{90mm} {\includegraphics{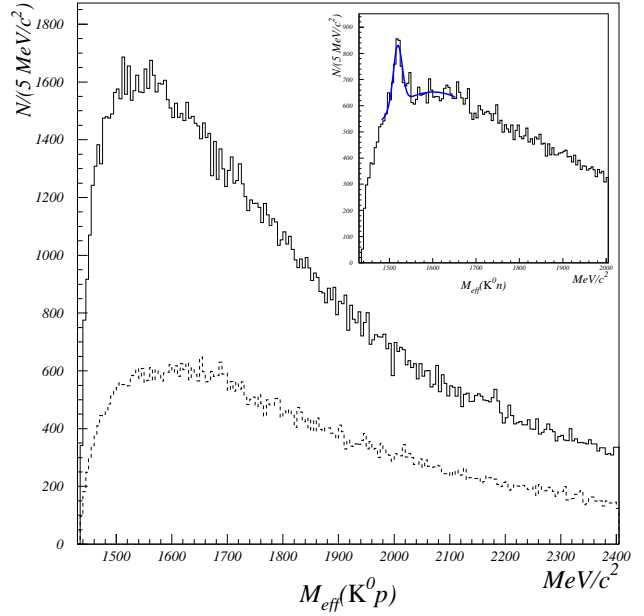}}
\caption{ The $(pK^0_s)$ invariant mass spectrum for positively
  charged particles considered as a proton obtained from RQMD
  simulation. Dashed histogram represents the $(pK^0_s)$ invariant
  mass spectrum with identified protons. Inset shows the $(nK^0_s)$
  invariant mass spectrum.}
\label{rqmd}
\end{figure}

The RQMD generator produces spectra including short lived
resonances. We used these spectra as input data for GEANT 3.21
package\cite{geant3} to achieve a set of final state particles
($K^0_s$ and others) which are detected by apparatus. This set
of particles was used to implement experimental cuts and to produce invariant
mass spectra. The $(nK^0_s)$ invariant mass spectrum clearly shows peak
of $\Lambda^*(1520)$ with 30\% enhancement over combinatorial
background. In contrast, the mass spectrum of the $pK^0_s$ system is
smooth and contains no fluctuations exceeding statistical errors on the
background(Fig.\ref{rqmd}).

\section{The $x_F\/$-distributions and cross sections.}

We performed two searches for $\Theta^+\/$ particle using SVD
'2002 data. Their independence is based primarily on using
$V^0\/$  candidate vertices in different regions. In both cases,
numerous checks were made on possible peak imitations. For the
kinematic reflections ($K^*(892)^+\rightarrow$ K$^{0}p\/$,
$\Sigma(1385)^+ \rightarrow K^{0}p$, etc.) and  mechanisms
with ghost tracks described in\cite{zavertyaev,longo}
null results were obtained.
A smooth shape of mixed-event background also excludes the possibility of
generating the sharp peak due to the applied cuts and the detector
acceptance. A narrow shape of the peak by itself makes it very
difficult to be generated by any kind of reflections or kinematic constraints
on the data.

It is impossible to determine the strangeness of this state in such
an inclusive reaction, however we did not observe any narrow structure in
$(\Lambda \pi^+)$ invariant mass spectrum in $1500\div1550~MeV/c^2$
mass region (Fig.\ref{kstar}b), expected for a particle with
the negative strangeness.

Our search for $\Theta^+$-particle is an inclusive experiment with a
significant background contribution. We have made an attempt to apply
a subtraction method to investigate $\Theta$ creation region in terms
of $x_F$.  An effective mass distribution around the peak was fitted
with a sum of background (B) and Gaussian functions, the latter having
a mean of $m_0\/$ and a standard deviation of $\sigma\/$. Background was
taken as a product of a threshold function and a second order
polynomial $P2$, \mbox{$B(m)=P2(m)(1-e^{-x_2(m-x_1)})$}, m standing for the
effective mass. All the fit parameters were given reasonable seeds but
no boundaries to prove a fit stability. The peak region was defined as
$m_0-2\sigma~<~m~<~m_0+2\sigma\/$. A number of effective
background events under the peak, $N(B_{peak})$, was evaluated by
integrating background function over the peak region. $x_F$ distributions
were taken separately for a peak and an out-of-peak regions(``wings'');
the latter ranged from 1.4 to 1.7 $GeV/c^2$ with a peak region cut out,
and a result was scaled to a $N(B_{peak})$. Assuming that the background
characteristics were uniform, we subtracted ``wings'' distribution from
the peak one. These operations were performed separately over the data
from both samples.

\begin{center}
  \begin{figure}[h]
    \vspace{35mm} {\includegraphics{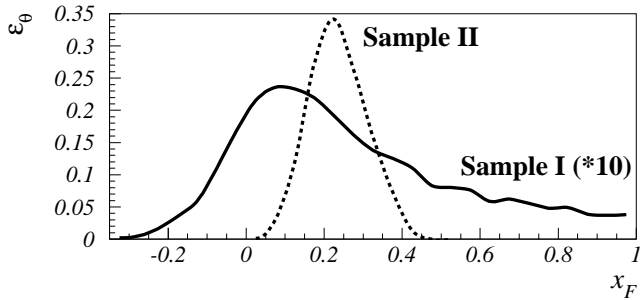}}
    \caption{Efficiencies of the $\Theta^+\/$ registration in $x_F$.
      Data for different targets are weighted using the total numbers of
      $K^0\/$ events. The result for a Sample I is scaled by 10.}
    \label{acc12}
  \end{figure}
\end{center}

Acceptance corrections were evaluated using a GEANT
Monte-Carlo\cite{geant3} for the $\Theta^+\/$ registration only.
We took into account a presence of $K^0\/$ decay in the selected region,
geometry acceptances, instrumental efficiencies and cuts used in the
analyses. We did not consider any other tracks, accompanying the
theta creation.
Theta particles were generated with a flat distribution over the
$-1<x_F<1$ region with a $<P_t>=0.52 GeV/c$, estimated from a Sample I
data and typical for the heavy particle creation.
A vertex of the decay was generated randomly over the measured beam spot
in X and Y, and Z was taken in the centers of the targets.
Reconstruction procedure was applied to the simulated
detector hits, taking into account detector efficiencies.
The overall result is shown in fig. \ref{acc12}.
The lower absolute value and a fall at  $x_F \to 1$ for the
Sample I are dominated by the limited region allowed for the $K^0_s$ decay,
and the narrow shape for the Sample II reflects the proton momentum
cut used.

\begin{figure}[h]
  \vspace{35mm} {\includegraphics{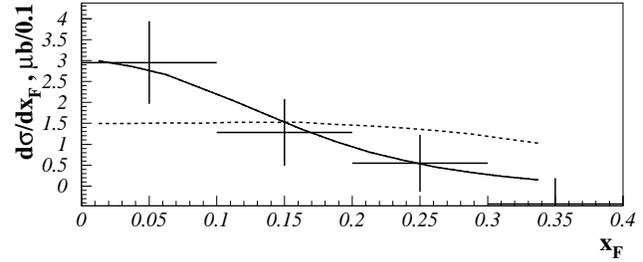}}
  \caption{$x_F$ distribution for the Sample I.
    Crosses are experimental data,
    curves stand for a quark fragmentation(solid line)
    and a sum of quark and diquark fragmentation(dotted line)
    in a Regge-based model\cite{baranov}.
    Y-axis: acceptance-corrected cross section over 0.1 in $x_F$.
    All the data are normalized to the 4.9~$\mu b$/nucleon for
    $0~<~x_F~<~0.3\/$ region}.
  \label{fig3}
\end{figure}

\begin{figure}[h]
  \vspace{35mm} {\includegraphics{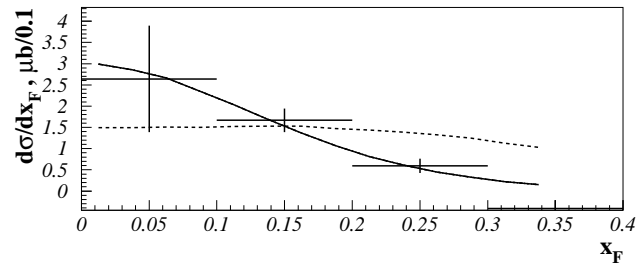}}
\caption{The same as in fig.\ref{fig3} for the Sample II.}
\label{fig4} \end{figure}

The results are shown in figs.\ref{fig3} and \ref{fig4}. We plot also
normalized curves of the predictions made in a Regge-based
model\cite{baranov}. In this model, an $x_F$ distribution comes from
the sum of quark fragmentation (bell-shaped around $x_F=0$) and diquark
one (seagull-shaped), each one with its proper weight. In
\cite{baranov} the weights were taken as 1:10, as coming from the
analysis of non-exotic baryons creation. Our data may indicate a
favoring to the quark fragmentation part of the model.

\begin{figure}[h]
  \vspace{35mm} {\includegraphics{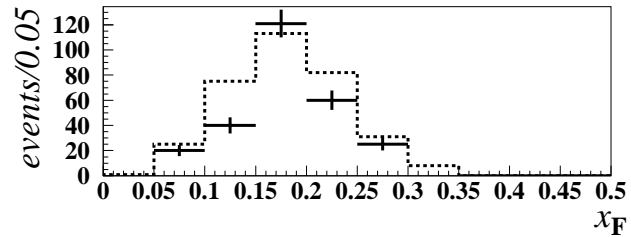}}
  \caption{A comparison of the Sample I (dotted histogram) to a
    Sample II (crosses) results.}
  \label{fig6}
\end{figure}

The Sample II has a specific
narrow acceptance in $x_F$ due to the proton momentum restrictions. We
projected the $x_F$-result of Sample I to a Sample II to check
the consistence of our observations. It was made by applying an
{\em inverse}  acceptance correction for Sample II to the result of
Sample I.
It is shown in fig.\ref{fig6} together with a raw (no acceptance
correction) data for a Sample II.
We found a plausible agreement of the distribution shapes
and a certain difference in a total number of events. The
latter makes a contribution to the cross section error calculations.

We have made an estimate of A-dependence in the $\Theta^+\/$ creation.
With SVD-2 vertex detector, Z-coordinate of primary vertex was reconstructed
to a precision of about 70 $\mu m$, so the events belonging to different
target planes were easily separated. For each material, a ratio was taken
 of the number of $\Theta^+\/$ events to the number of all the $K^0_s$ events.
 Weighting it to a $C^{12}$ value (table \ref{table_eff1}), we found no
differences in the material dependence.
Assuming that the observed $K^0\/$ events exhibit a
$A^{0.71}\/$-dependence, we estimate that $\Theta^+\/$ creation follows
the same law (within errors).

\begin{table}
  \caption{A-dependence. The fractions of $\Theta^+\/$-events
    in the $K^0$ selection for different materials,
    weighted to a $C^{12}$ result.}
  \begin{center}
    \begin{tabular}{|l|r|}
      \hline
      Target & Relative cross section\\
      \hline
      $C^{12}$ &   $1.0 \pm 0.22$ \\
      \hline
      $Si^{28}$ &  $0.87 \pm 0.29$  \\
      \hline
      $Pb^{207}$ & $1.09 \pm 0.32$ \\
      \hline
    \end{tabular}
  \end{center}
  \label{table_eff1}
\end{table}

The cross section of $\Theta^+$ creation in pN interactions for $x_F>0$
was evaluated from Sample I data (as having a better acceptance at $x_F\to 0$):
\begin{eqnarray}
  \sigma_{\Theta pN}\cdot Br(\Theta^+\rightarrow pK^0) =
  \frac{\sigma_{pN}\cdot \epsilon_{corr}}{0.34}
  \frac{N_{\Theta}}{N_{pN} \cdot \epsilon_\Theta} \nonumber \\
= 4.9 \pm 1.0(stat.) \pm 1.5(syst.)~ \mu b/nucleon , \nonumber
\end{eqnarray}
where $\sigma_{pN}=31.14~\mu b~$ is the inelastic proton-proton cross
section\cite{LA0},  0.34 is the PDG\cite{pdg} value for the
Br($K^0\to\pi^+\pi^-$),
$N_{\Theta}$ and $N_{pN}$ are the numbers of events with a $\Theta^+$ particle
and all the events respectively and $\epsilon_\Theta\/$ is the efficiency
of $\Theta$ registration. A luminousity correction factor
$\epsilon_{corr}=0.84$ was evaluated from the following considerations.
A trigger used for the physics events was developed initially as a Level I
trigger, designed to be sensitive to all the inelastic $pA\/$-events in the
target planes. A comparison to the ``beam'' trigger events showed that in our
samples a registration of the primary vertices with a charged particles
multiplicity $N_{ch}\leq 3\/$ is suppressed. We estimate the fraction of total
inelastic events accepted by trigger as 0.63.
Next, in a Sample I we cut out the events with $N_{ch}\geq 8\/$
accounting for 0.4 of all the registered events.
Assuming that $\Theta^+$-creation is 1) suppressed in the low $N_{ch}$ region
and 2) does not depend on the
multiplicity in the high $N_{ch}$ region, we derived a combined scaling
coefficient for the $\Theta^+$ cross section as
$\epsilon_{corr}=0.84$. As the better estimate of the cross section
dependence on the multiplicity was not available, the resulting systematic
error is rather large. With that correction, our integrated
luminousity was $2.3\cdot 10^{33} cm^{-2}$. It was implicitly assumed that
$\sigma_{\Theta}\/$ in pA interactions also scales as $A^{0.71}\/$.
This cross section value agrees within errors to the result published in
\cite{svdichep06}, but differs from our earliest
estimate\cite{svd4}. The main reason is an unexpected form of $x_F$
distribution, assumed flat in our first publications. Taking into
account our situation of using inclusive data, and undiscovered yet
mechanism of theta particle creation, we believe that the cross section
value is subject to future investigations.

\begin{figure} [ht]
  \vspace{45mm}
	 {\includegraphics{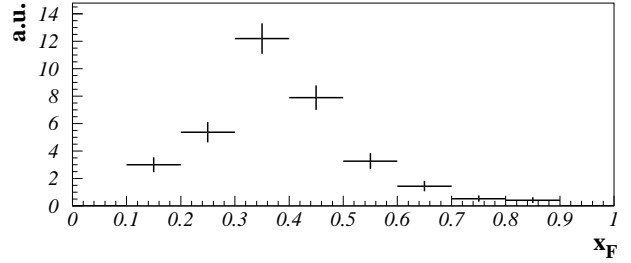}
	 }
  \caption{An $x_F$ distribution for a $\Lambda (1520)$.}
  \label{lam1520xf}
\end{figure}

As positive as negative results of the theta particle searches were
very often presented as the ratio of theta to $\Lambda (1520)$ cross
sections. We estimate the latter value as $147 \pm 35 ~\mu b/nucleon$,
that agrees with the result obtained by EXCHARM
collaboration\cite{excharm} and gives
the ratio of 0.07 assuming that the $\Theta^+ \to pK^0$ and
$\Theta^+ \to nK^+$
decays have equal probabilities. However these two
particles may have quite different creation
mechanisms. For example, extracting $x_F$ for $\Lambda (1520)$ by the same
method as described above resulted in much more forward-oriented distribution
(fig.\ref{lam1520xf}), than for a theta particle. It weakens the
reasons of making this comparison. We may suppose that a best
way to present such a ratio would be to accompany it with
corresponding acceptances and $x_F$ distributions.

\section {Summary and Conclusions.}

The inclusive reaction $p A \rightarrow pK^0_s + X$ was studied at
IHEP accelerator with $70\ GeV/c$ proton beam using SVD-2 detector. Two
different samples of $K^0_s$, statistically independent and belonging
to different phase space regions, were used in the analyses and a
narrow baryon resonance with the mass $M=1523\pm 2(stat.)\pm
3(syst.)\ MeV/c^2$ was observed in both samples of the data.
We observed the total of s=392 signal events
over b=1990 of background. Using additional track quality cuts we
obtained $\Gamma < 14\ MeV/c^2$ on 95\% C.L.  The statistical
significance of the peak can be estimated to be (for different
estimators):\\
s/$\sqrt{b}$ = 8.7, s/$\sqrt{s + b}$ = 8.0, s/$\sqrt{s + 2b}$ = 5.9.

\noindent
The new analysis and larger statistics confirm our previous
result\cite{svd4}. The resonance observed is not a statistical
fluctuation neither induced by background processes. We plan to
continue our investigations of the creation mechanisms of
$pK^0_s$ resonance, momentum and angular dependencies and cross
sections. The cross-section estimate for $x_F > 0$
($\sigma \cdot BR(\Theta^+ \rightarrow pK^0) = 4.9 \ \mu b)$ was
obtained. $x_F$ distribution was found to peak at zero with
\mbox{$<|x_F|> \approx 0.1$}, that agrees
qualitatively to a Regge-based model\cite{baranov}. While in agreement
with evidences of $\Theta^+$ observation, there is no direct
contradiction to null results in hadron-hadron fixed target
experiments (e.g. \cite{sphinx}), mainly due to different acceptances
at $x_F \approx 0$.

We acknowledge the efforts of IHEP management in providing us with the
accelerator time and infrastructure support.
We wish to thank Ya.~Azimov, S.~Baranov, L.~Gladilin and D.~Melikhov
for useful comments and suggestions. We are grateful to Dr. V. Voevodin and
Scientific Research Computing Center of Moscow State University for granting
an access to the high-perfomance supercomputing cluster.

This work was supported by Russian Foundation for Basic Research (N
03-02-16894, 06-02-16954, 06-02-81010-Bel\_a),
the Program "Universities of Russia" (N UR 02-02-505),
Russian Foundation for leading scientific schools (N 1685-2003-02,
8122-2006-02) and by the contract with Russian Ministry of Industry,
Science and Technology (Gos\-kon\-t\-rakt N 40-032-11-34) in the part of the
development and creation of the vertex detector.

\end{document}